\documentclass[3p]{elsarticle}

\usepackage{hyperref}

\journal{Elsevier}

\begin{document}

\begin{frontmatter}

\title{Influence of experimental conditions on localized surface plasmon resonances measurement by electron energy loss spectroscopy}

\author[ceitec]{Michal Hor\'{a}k\corref{correspondingauthor}}
\cortext[correspondingauthor]{Corresponding author}
\ead{Michal.Horak2@ceitec.vutbr.cz}

\author[ceitec,ufi]{Tom\'{a}\v{s} \v{S}ikola}

\address[ceitec]{CEITEC -- Central European Institute of Technology, Brno University of Technology, Purky\v{n}ova 123, 612 00 Brno, Czech Republic}

\address[ufi]{Institute of Physical Engineering, Brno University of Technology, Technick\'{a} 2, 616 69 Brno, Czech Republic}

\begin{abstract}
Scanning transmission electron microscopy (STEM) combined with electron energy loss spectroscopy (EELS) has become a standard technique to map localized surface plasmon resonances with a nanometer spatial and a sufficient energy resolution over the last 15 years. However, no experimental work discussing the influence of experimental conditions during the measurement has been published up to now. We present an experimental study of the influence of the primary beam energy and the collection semi-angle on the plasmon resonances measurement by STEM-EELS. To explore the influence of these two experimental parameters we study a series of gold rods and gold bow-tie and diabolo antennas. We discuss the impact on experimental characteristics which are important for successful detection of the plasmon peak in EELS, namely: the intensity of plasmonic signal, the signal to background ratio, and the signal to zero-loss peak ratio. We show that the best results are obtained using a medium primary beam energy, in our case 120\,keV, and an arbitrary collection semi-angle, as it is not a critical parameter at this primary beam energy. Our instructive overview will help microscopists in the field of plasmonics to arrange their experiments.
\end{abstract}


\end{frontmatter}

\section{Introduction}

Mapping of localized surface plasmon resonances (LSPR), collective oscillations of free electrons in metallic nanostructures coupled to the local electromagnetic field \cite{novotny2011,schuller2010,kelly2003}, with high spatial and energy resolution is necessary to understand their origin and properties. The best spatial, energy, and time resolution is achieved by electron beam spectroscopy \cite{abajo2010,losquin2016}. Scanning transmission electron microscopy (STEM) combined with electron energy loss spectroscopy (EELS) has become a standard technique to map LSPR with a nanometer spatial and 10\,meV to 100\,meV energy resolution over the last 15 years \cite{nelayah2007,kociak2014,colliex2016,wu2018,tizei2018}.

Despite that many works dealing with EELS measurement of LSPR have been published, there is no experimental work discussing the experimental conditions during the measurement. Such an instructive overview has been missing in literature making it especially difficult for the beginners in the field. Experimental parameters of the STEM-EELS measurement, such as the primary beam energy, the convergence semi-angle and the collection semi-angle, may influence obtained results. Many publications include incomplete information about these parameters or miss them completely. Moreover, even if they are in the literature completely included, one can find several different approaches. Some of them are summarized in Table \ref{Tab1}. Primary beam energy is used in the range from 60\,keV to 300\,keV. At 60\,keV and 80\,keV, microscopists use the collection semi-angle smaller than the convergence semi-angle \cite{schaffer2010,tizei2018,bitton2020}. At all higher primary beam energies, the collection semi-angle similar to the convergence semi-angle is used quite often \cite{nelayah2007,raza2013,zhou2014a,krapek2020,koh2009,koh2011}. And finally at 300\,keV, the collection semi-angle larger than the convergence semi-angle is sometimes used \cite{haberfehlner2015,horak2018,horak2019,krapek2015}. All presented experimental results are clear with no suspicion of any bad STEM-EELS measurement. Consequently, there is no clear conclusion which experimental conditions are optimal.

\begin{table}[h]
\caption{Experimental conditions of STEM-EELS measurements of LSPR in the literature.}
\label{Tab1}
\centering
\begin{tabular}[h]{|c|c|c|c|}
\hline
Primary & Convergence & Collection & \\
beam energy & semi-angle & semi-angle & Reference \\
\hline
60\,keV & 34\,mrad & 10\,mrad & \cite{tizei2018}\\
80\,keV & 11.4\,mrad & 2\,mrad & \cite{schaffer2010}\\
80\,keV & 34\,mrad & 18\,mrad & \cite{bitton2020}\\
100\,keV & 7.5\,mrad & 6\,mrad & \cite{nelayah2007}\\
120\,keV & 15\,mrad & 17\,mrad & \cite{raza2013}\\
200\,keV & 9\,mrad & 12\,mrad & \cite{cherqui2016}\\
300\,keV & 10\,mrad & 6.6\,mrad & \cite{ligmajer2019}\\
300\,keV & 13\,mrad & 13\,mrad & \cite{zhou2014a}\\
300\,keV & 10\,mrad & 10.4\,mrad & \cite{krapek2020}\\
300\,keV & 15.1\,mrad & 17.4\,mrad & \cite{koh2009,koh2011}\\
300\,keV & 15\,mrad & 20.5\,mrad & \cite{haberfehlner2015}\\
300\,keV & 10\,mrad & 20.5\,mrad & \cite{horak2018,horak2019}\\
300\,keV & 9.2\,mrad & 20.7\,mrad & \cite{krapek2015}\\
\hline
\end{tabular}
\end{table}

In our contribution, we present the influence of these experimental conditions, namely the primary beam energy and the convergence and the collection semi-angle, on LSPR measurement by STEM-EELS while our figure of merit is the intensity of plasmonic signal, the signal to background ratio, and the signal to zero-loss peak ratio considering a limited dynamic range of the spectrometer camera. Our goal is to find the optimal experimental conditions and give an instructive overview which would help microscopists in the field of plasmonics to arrange their experiments.

\section{Theory}

EELS utilizes an electron beam that interacts with the metallic nanoparticle and excites the LSPR. In consequence, the energy of some electrons decreases by the characteristic energy of the LSPR. As the result, we observe a peak in the electron energy loss spectrum at the respect energy loss. Spatially-resolved EELS further provides (relative) intensity of the near electric field of a LSPR projected to the trajectory of the electron beam.

The field induced in the specimen by the probing electron acts back on the electron and decreases its energy with the loss probability density $\Gamma^\mathrm{EELS}(\vec{r_\mathrm{e}}(t),\omega)$ reading \cite{abajo2010}
\begin{equation}
\label{eq1}
\Gamma^\mathrm{EELS}=\frac{e}{\pi \hbar \omega}\int\mathrm{d}t \Re\left\lbrace\mathrm{e}^{-i\omega t}\vec{v}\cdot \vec{E}_\mathrm{el}^\mathrm{ind}(\vec{r_\mathrm{e}}(t),\omega)\right\rbrace,
\end{equation}
where $\hbar \omega$ is the loss energy, $\vec{r_\mathrm{e}}(t)$ represents the electron trajectory, $\vec{v}$ is the electron velocity, and $\vec{E}_\mathrm{el}^\mathrm{ind}(\vec{r_\mathrm{e}}(t),\omega)$ denotes the induced parts of the electric field by the fast electron (for example, the field of LSP). It is calculated by removing the free-space solution of the electric field from the total computed electric field. Correspondence between the loss probability and the electromagnetic local density of states projected along the electron beam trajectory has been discussed \cite{kociak2014,abajo2008,hohenester2009}.

Equation (\ref{eq1}) tells us that the loss probability density measured by EELS depends on the electron velocity which is directly related to the primary beam energy in the STEM. However, electron losses are contributed not only by LSPR-related losses proportional to the out-of-plane near field but also by material-related losses proportional to the relative thickness of the sample and relativistic effects like the \v{C}erenkov radiation \cite{horak2015} dependent on the speed of the electron. To separate the contribution of LSPR losses from other losses, background subtraction is usually performed. The background is the EEL spectrum recorded on a pure substrate far away from the plasmonic structure, so it does not contain any LSPR-related signal. All losses have some angular distribution which is very narrow at the energies in units of electronvolts \cite{egerton}. As the angular distribution of all losses is not exactly the same, the choice of a proper collection semi-angle may play an important role in the STEM-EELS experiment. Therefore, the primary beam energy and the collection semi-angle are important experimental parameters. 

\begin{figure}[h!]
\centering
\includegraphics[width=8cm]{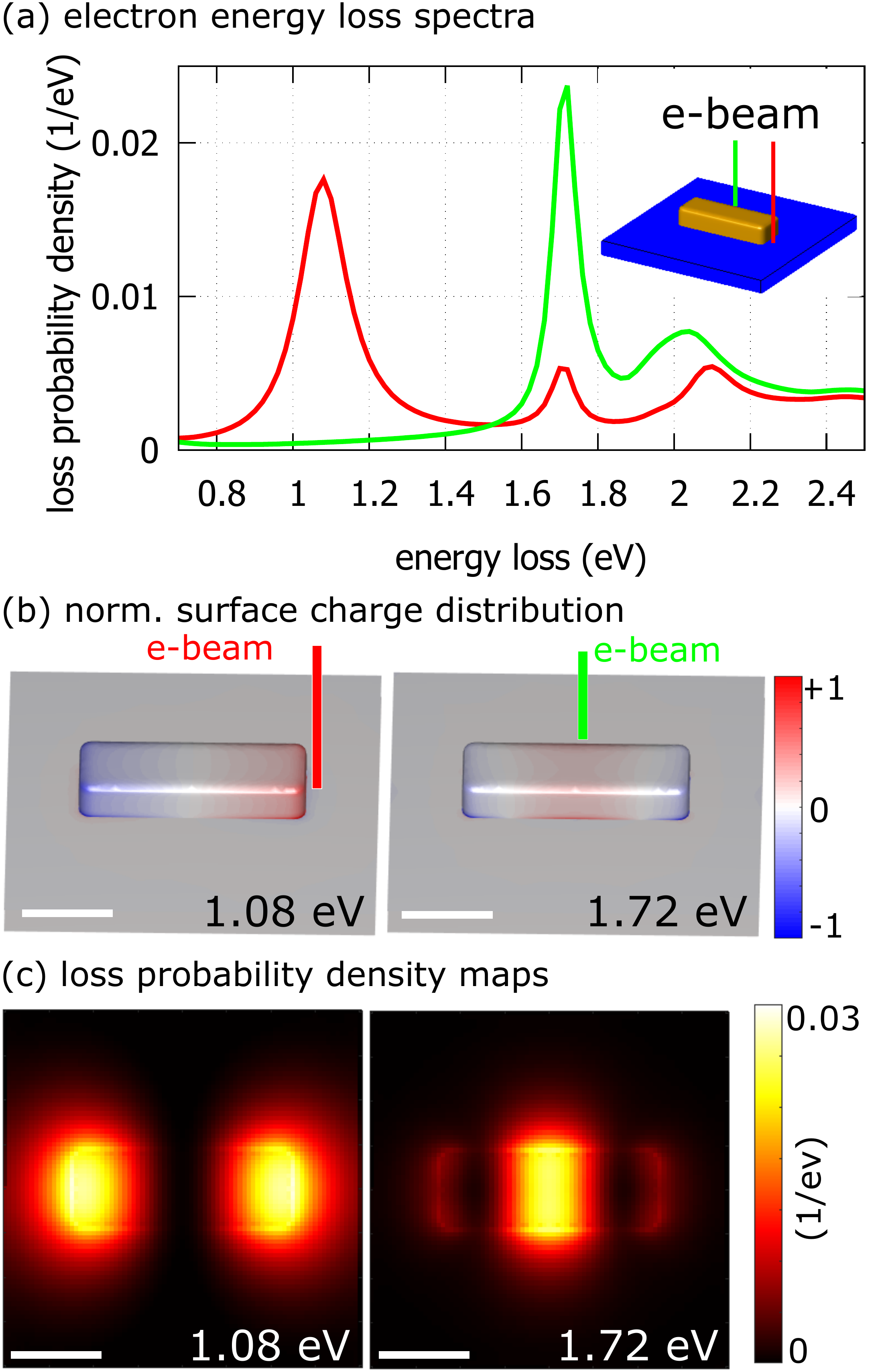}
\caption{Numerical simulation of LSPR in a gold rod (length 240\,nm, width 80\,nm, height 30\,nm) on a 30\,nm thick SiN$_\mathrm{x}$ membrane: EEL spectra (a) for two positions (5\,nm outside the rod at the middle of its shorter and longer edge) of the 300\,keV electron beam followed by surface charge distribution (b) and loss probability maps (c) for the longitudinal dipole mode at 1.08\,eV and the longitudinal quadrupole mode at 1.72\,eV. The scalebar in (b) and (c) is 100\,nm long.}
\label{Fig1}
\end{figure}

We have studied a series of gold rods to explore the influence of these two experimental parameters. Plasmonic nanorods represent very simple objects which have been studied many times by EELS \cite{bryant2008,zhou2014,bosman2014}. Our gold rods with the length of 240\,nm, the width of 80\,nm, and the height of 30\,nm are situated on a 30\,nm thick silicon nitride membrane. First, we have described the structure theoretically. Numerical simulations were performed with the MNPBEM toolbox \cite{waxenegger2015} which relies on the boundary element method (BEM) approach \citep{abajo1998,abajo2002}. The dielectric function of single-crystal gold was taken from Olmon et al. \cite{olmon2012} and the dielectric function of the silicon nitride membrane was set to 4. The rod supports two main LSPR modes -- the longitudinal dipole mode at 1.08\,eV and the longitudinal quadrupole mode at 1.72\,eV. Their spectral and spatial characteristics are introduced in Figure \ref{Fig1}.

Finally, to ensure that obtained results are not valid exclusively for rods, we have included into our study several bow-tie and diabolo antennas, which have been recently studied theoretically \cite{hrton2020} and experimentally \cite{krapek2020}. All structures were of a similar size: wing angle of 90$^\circ$ and wing length around 250 nm. We have studied the transverse dipole mode in both types of antennas (referred to as TD mode) and the longitudinal dipole mode in the bow-tie or the longitudinal dipole antibonding mode in the diabolo, respectively (both referred as LD mode in the following for simplicity), as they are well defined by their spatial distribution and are the same for bow-tie and diabolo antennas. These antennas support the TD mode (with the maximal loss probability located at four outer corners of the antenna) around 0.8\,eV and the LD mode (with the maximal loss probability located in the gap/bridge of the antenna) around 1.2\,eV \citep{krapek2020}.

\section{Experimental methods}

Gold rods with the length of 240\,nm, width of 80\,nm, and height of 30\,nm and bow-tie and diabolo antennas with the wing angle of 90$^\circ$ and wing length of 250 nm were prepared using focused ion beam (FIB) lithography by gallium ions (using dual beam FEI Helios) \cite{horak2018} of a 30\,nm thick gold layer deposited by magnetron sputtering (using Leica coater EM ACE600) directly on a 30\,nm thick silicon nitride membrane (purchased from Agar Scientific). We did not use any adhesion layer as it affects and degrades the LSPR \cite{habteyes2012,madsen2017}. The sample was plasma cleaned in argon-oxygen plasma for 20 seconds (using plasma cleaner Fishione 1020) to prevent the carbon contamination during the STEM-EELS measurement \cite{horak2018}.

EELS measurements were performed with TEM FEI Titan using a double focusing Wien filter monochromator equipped with GIF Quantum spectrometer operated in a monochromated scanning regime at the primary beam energy of 60\,keV, 120\,keV, and 300\,keV, the beam energies to which our microscope is tuned. Beam current was set around 0.2 nA and the full-width at half-maximum of the zero-loss peak (ZLP) was in range from 0.1\,eV to 0.15\,eV. We set the convergence semi-angle to 10\,mrad while the collection semi-angle was changed from 1.3\,mrad to 20.5\,mrad by changing the camera length of the microscope in the STEM mode. The dispersion of the spectrometer was 0.01\,eV/pixel. We recorded spectrum images with the pixel size of 1\,nm or 2\,nm. Every pixel consists of 1 EEL spectrum whose acquisition time was adjusted to use the maximal intensity range of CCD camera in the spectrometer and avoid its overexposure.

\begin{figure}[h!]
\centering
\includegraphics[width=8cm]{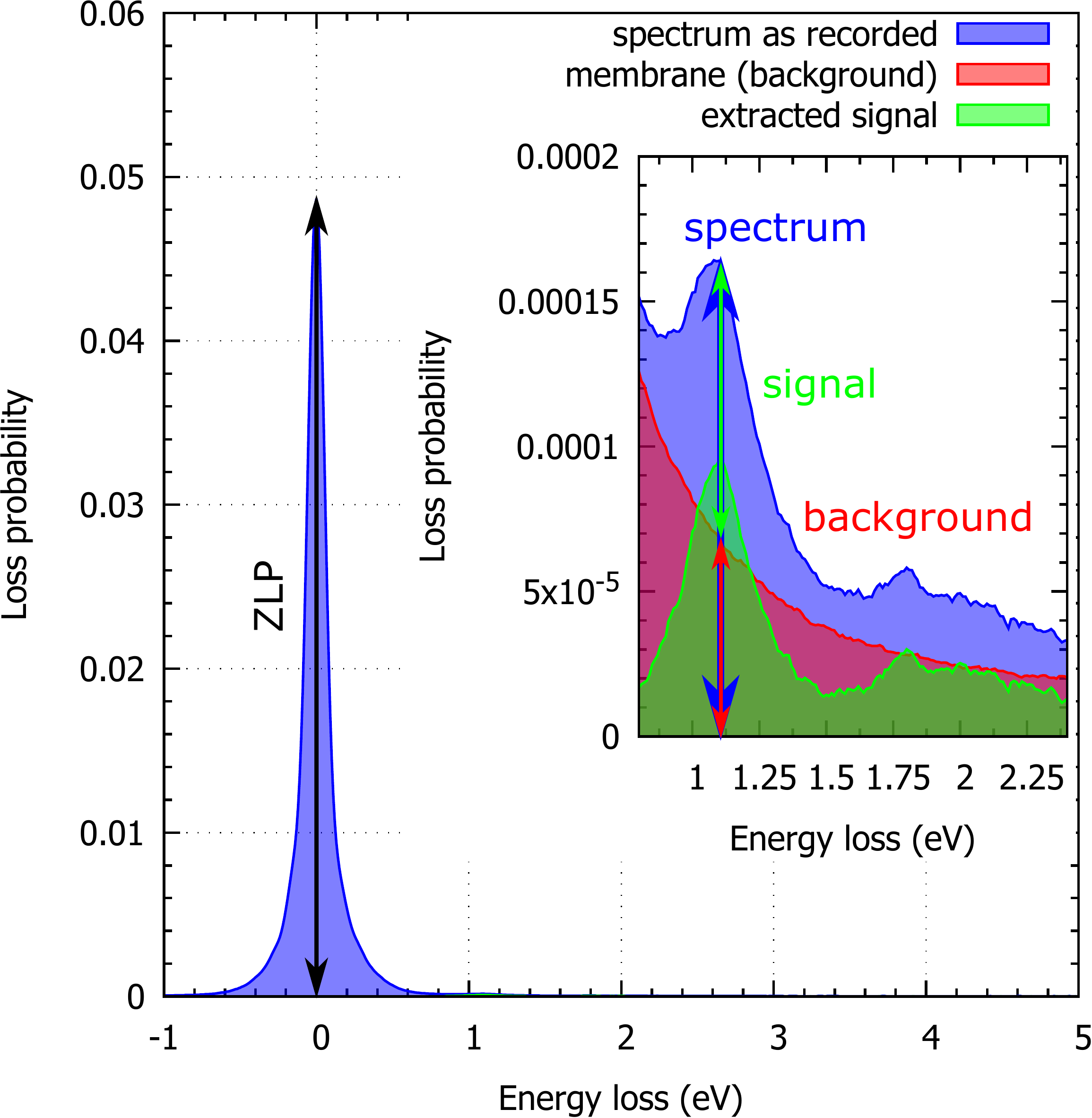}
\caption{Typical raw low-loss EEL spectrum (blue) decomposed into background represented by the spectrum of pure SiN$_\mathrm{x}$ membrane far away from the rod (red) and the signal corresponding to the LSPR (green). We define the following quantities: ZLP as the loss probability at 0\,eV, spectrum as the loss probability at the energy of LSPR, signal as the loss probability at the energy of LSPR corresponding to extracted signal, and background as the loss probability at the energy of LSPR corresponding to the subtracted background.}
\label{Fig2}
\end{figure}

EEL spectra were integrated over rectangular areas consisting of tens of pixels in the places of interest where the respect LSPR is significant. Finally, the spectrum was divided by the integral intensity of the ZLP (the energy window for integration was from -1\,eV to +1\,eV) so the counts were transformed to a quantity proportional to the loss probability at 0.01\,eV intervals (referred as the loss probability in the following for simplicity). To separate the raw low-loss EEL spectrum (referred to as the spectrum) to the contribution of LSPR losses (referred to as the signal) from other losses (referred to as the background) we have subtracted the EEL spectrum recorded on a pure membrane far away from the gold rod (Figure \ref{Fig2}).

EEL maps were calculated by dividing the map of integrated intensity at the plasmon peak energy with the energy window of 0.1\,eV by the map of the integral intensity of the ZLP (energy window from -1\,eV to +1\,eV).

\section{Results}

We introduce several important experimental characteristics based on definitions in the previous section and in Figure \ref{Fig2}. To detect the peak corresponding to a LSPR as easily as possible one needs to maximize the signal itself (in principle the signal to noise ratio), to maximize the signal to background ratio, and to maximize the signal to ZLP ratio as the spectrometer camera has a limited dynamic range. The influence of the primary beam energy and the collection semi-angle on these characteristics is presented in the two following sections for rods. Finally, to ensure that obtained results are not valid exclusively for rods, we have included to our study several bow-tie and diabolo antennas, too.

\subsection{Influence of the primary beam energy on LSPR signal from rods}

In the first experiment we take a series of 3 rods and do the STEM-EELS measurement at the primary beam energy of 300\,keV, 120\,keV, and 60\,keV while keeping the the collection semi-angle constant at 20.5\,mrad.

\begin{figure}[h!]
\centering
\includegraphics[width=8cm]{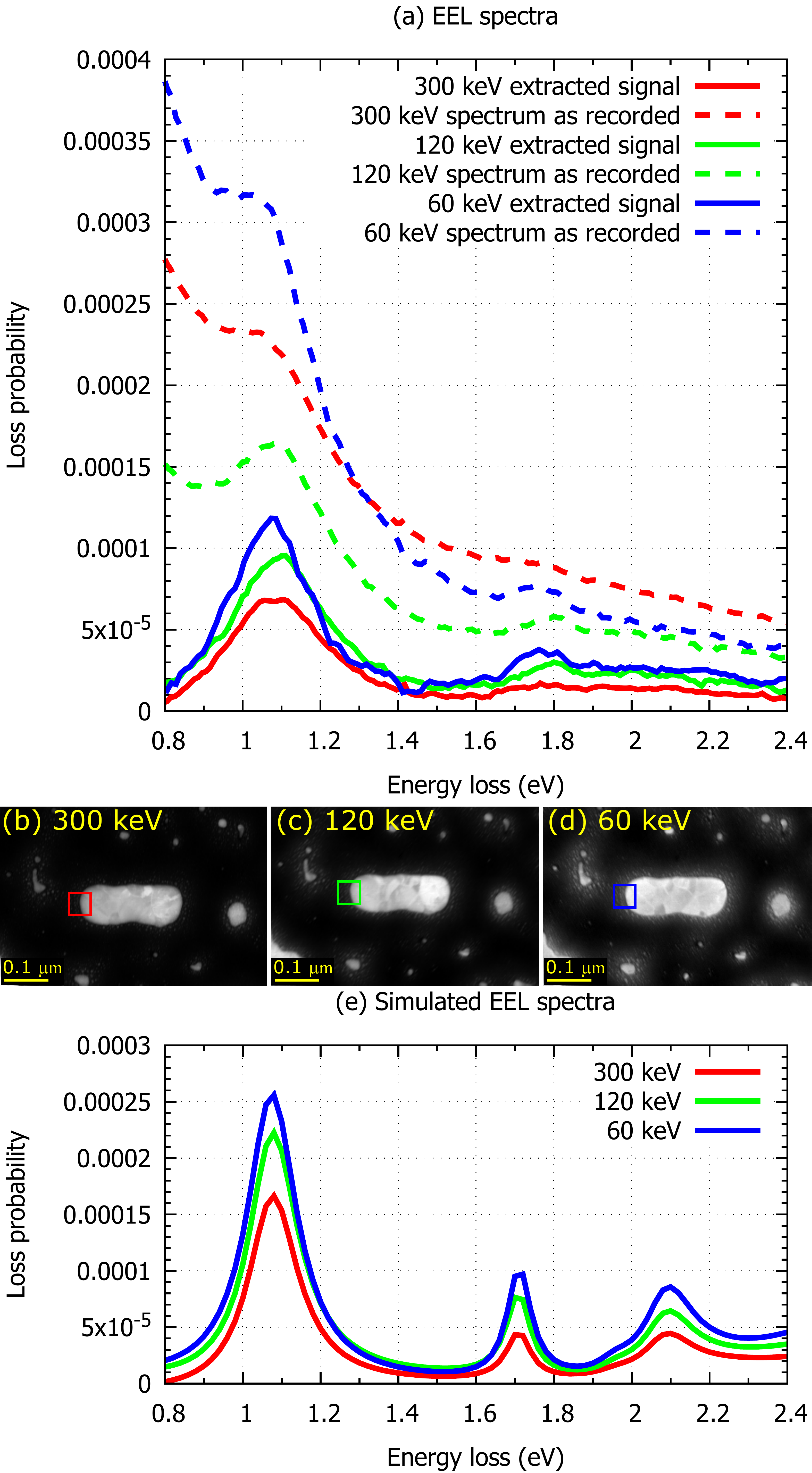}
\caption{EEL spectra of the same rod at different beam energies: (a) measured raw EEL spectra and extracted signal; (b-d) STEM annular dark field (ADF) images of the rod with marked area for integration of EEL spectra in (a) recorded during STEM-EELS mapping at 300\,keV (b), 120\,keV (c), and 60\,keV (d); (e) simulated EEL spectra by BEM. Electron beam was situated 5\,nm outside the rod at the middle of its shorter edge and calculated loss probability density was recalculated to loss probability at 0.01\,eV energy intervals (corresponding to the dispersion of the spectrometer in the experiment).}
\label{Fig3}
\end{figure}

Figure \ref{Fig3}(a) shows EEL spectra of one rod integrated over the same selected area recorded at different beam energies. The areas of integration are marked by colored squares in STEM annular dark field (ADF) micrographs of the rod in Figure \ref{Fig3}(b-d). We clearly see that the signal corresponding to the LSPR is the strongest for the 60\,keV electron beam and the weakest for the 300\,keV electron beam. This experimental finding is in a perfect agreement with the theory represented by BEM simulation of EEL spectra shown in Figure \ref{Fig3}(e). Moreover, if we consider the second peak in experimental EEL spectra in Figure \ref{Fig3}(a) at 1.76\,eV corresponding to the longitudinal quadrupole mode, we see that this mode is, in principle, not detected when using 300\,keV electron beam, but is clearly visible when using 60\,keV or 120\,keV electron beam. Consequently, lower beam energies are better for observation of weaker plasmon modes. Based on this observation, one can say that the lower the energy of primary electron beam, the better.

If we consider measured raw EEL spectra [dashed lines in Figure \ref{Fig3}(a)], we clearly see that the peak at 1.08\,eV corresponding to the longitudinal dipole mode is the most noticeable when using 120\,keV electron beam. In the case of 300\,keV electron beam the background is enhanced by relativistic effects like the \v{C}erenkov radiation as the speed of the 300\,keV electron is higher than the speed of the light in the silicon nitride membrane with the refractive index around 2 \cite{philipp1973}. Therefore, the \v{C}erenkov limit for silicon nitride is around 80\,keV and the probability of \v{C}erenkov photon excitation per unit path length of the electron inside the silicon nitride specimen equals to 0.3 when the beam energy is around 120\,keV \cite{horak2015}. Considering this, the primary beam energy should not exceed the theoretical \v{C}erenkov limit for the membrane too much and the probability of \v{C}erenkov photon excitation per unit path length should be below 0.4, which was introduced as the experimental limit for daily use \cite{msp2008}. On the other hand, the raw EEL spectra measured with a 60\,keV electron beam has the highest background in the lower energy loss region. This is caused by a higher probability of scattering events as the inelastic mean free path (IMFP) of electrons in the sample is smaller \cite{egerton} or, in other words, the relative thickness of the sample is larger. Consequently, the optimal primary beam energy should not be too low to keep the relative thickness of the supporting membrane as low as possible, which can be achieved either by increasing the beam energy until the relativistic effects become too strong.

\begin{figure}[h!]
\centering
\includegraphics[width=10cm]{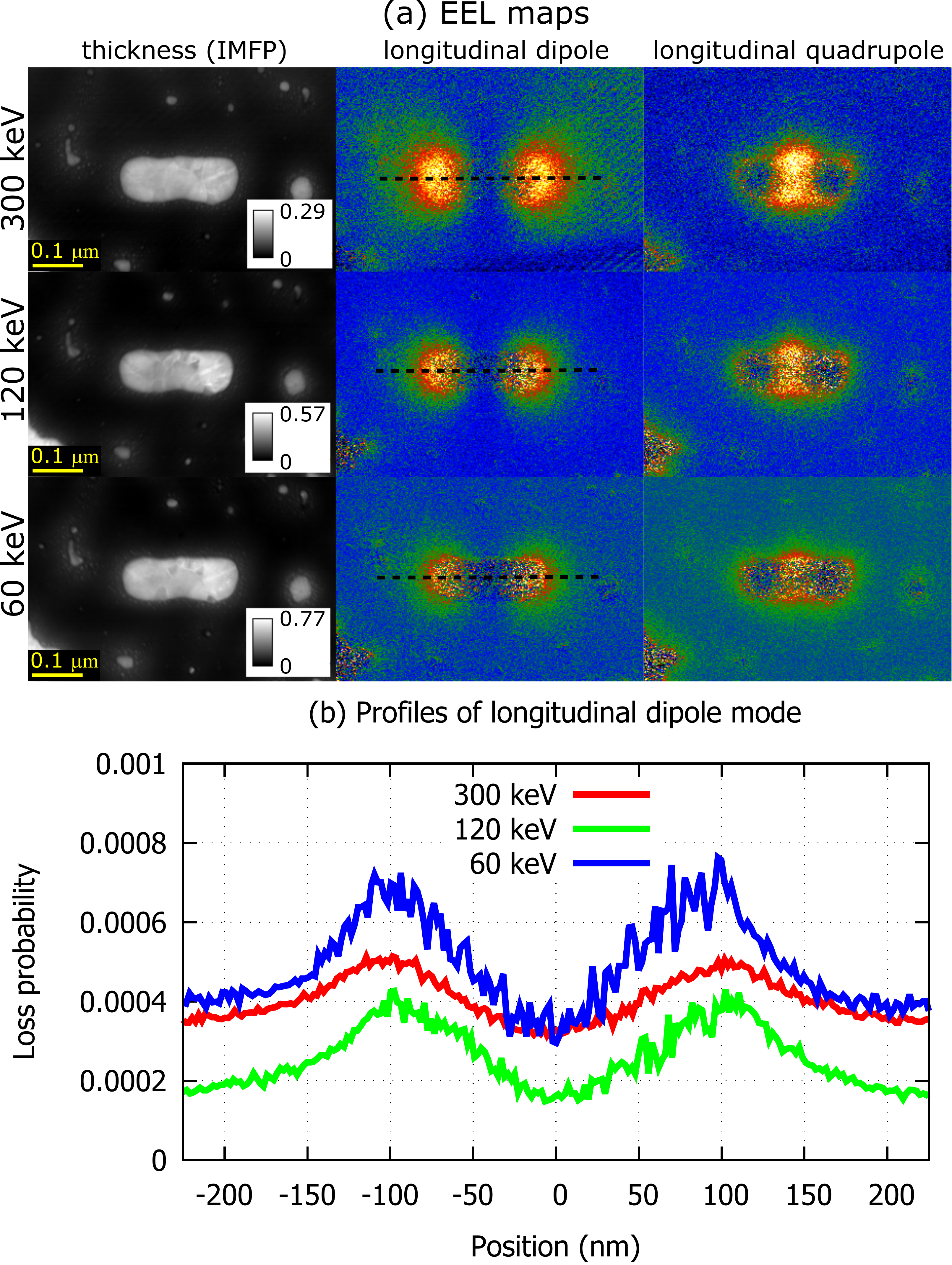}
\caption{(a) Maps measured by EELS: thickness maps [the thickness of gold is evaluated in units of inelastic mean free path (IMFP) of electrons in gold] of the rod and EEL maps of the longitudinal dipole mode at 1.08\,eV and the longitudinal quadrupole mode at 1.76\,eV recorded at different primary beam energies (300\,keV, 120\,keV, and 60\,keV). In the EEL maps the temperature color scale is used, i.e. yellow corresponds to maximal values and blue corresponds to minimal values. (b) Line profiles of the EEL maps of the longitudinal dipole mode along the dashed black lines in (a). The zero position corresponds to the middle of the rod.}
\label{Fig4}
\end{figure}

Figure \ref{Fig4}(a) shows relative thickness maps (i.e. the thickness of gold is evaluated in units of inelastic mean free path of electrons in gold) and EEL maps of the longitudinal dipole mode and the longitudinal quadrupole mode in the same rod recorded at different primary beam energies. First, we will inspect the relative thickness maps of the rods. The maximal relative thickness of the gold in the area of the respect spectrum image reads 0.29 for 300\,keV, 0.57 for 120\,keV, and 0.77 for 60\,keV. The relative thickness is thus $2.66\times$ ($1.97\times$) larger for 60\,keV (120\,keV) primary electron beam energy in comparison to the 300\,keV electrons. The increased relative thickness of the gold plays an important role when the EEL map is recorded. As the whole map is recorded with the same acquisition time per spectrum, spectra recorded at positions in the gold will have generally much lower signal in the energy range of LSPR as more electrons will scatter inelastically. Consequently, for lower beam energies one can expect more noisy EEL maps of LSPR.

Now we focus on EEL maps in Figure \ref{Fig4}(a). We easily observe two phenomena. First, the EEL maps recorded at 60\,keV are rather noisy. Second, the spatial distribution of the LSPR seems to be more confined when measuring with lower beam energies. However, the latter is just an optical illusion caused by the color scale. Inspecting the line profiles in Figure \ref{Fig4}(b) we clearly see that the spatial distribution of the LSPR is similar for all three primary beam energies. Consequently, the only difference in the EEL maps is in the signal to noise ratio. Therefore, the optimal primary beam energy should be high enough to measure the signal at positions in the metal with a good signal to noise ratio. This is the easiest solution as it is usually not possible to decrease the thickness of the metal or to adjust optimal exposure time for every pixel. The former one would change the studied object. The latter one would significantly prolonged the measuring time as some auto-exposure procedure needs to be applied at each pixel of the spectrum image.

\begin{figure}[h!]
\centering
\includegraphics[width=8cm]{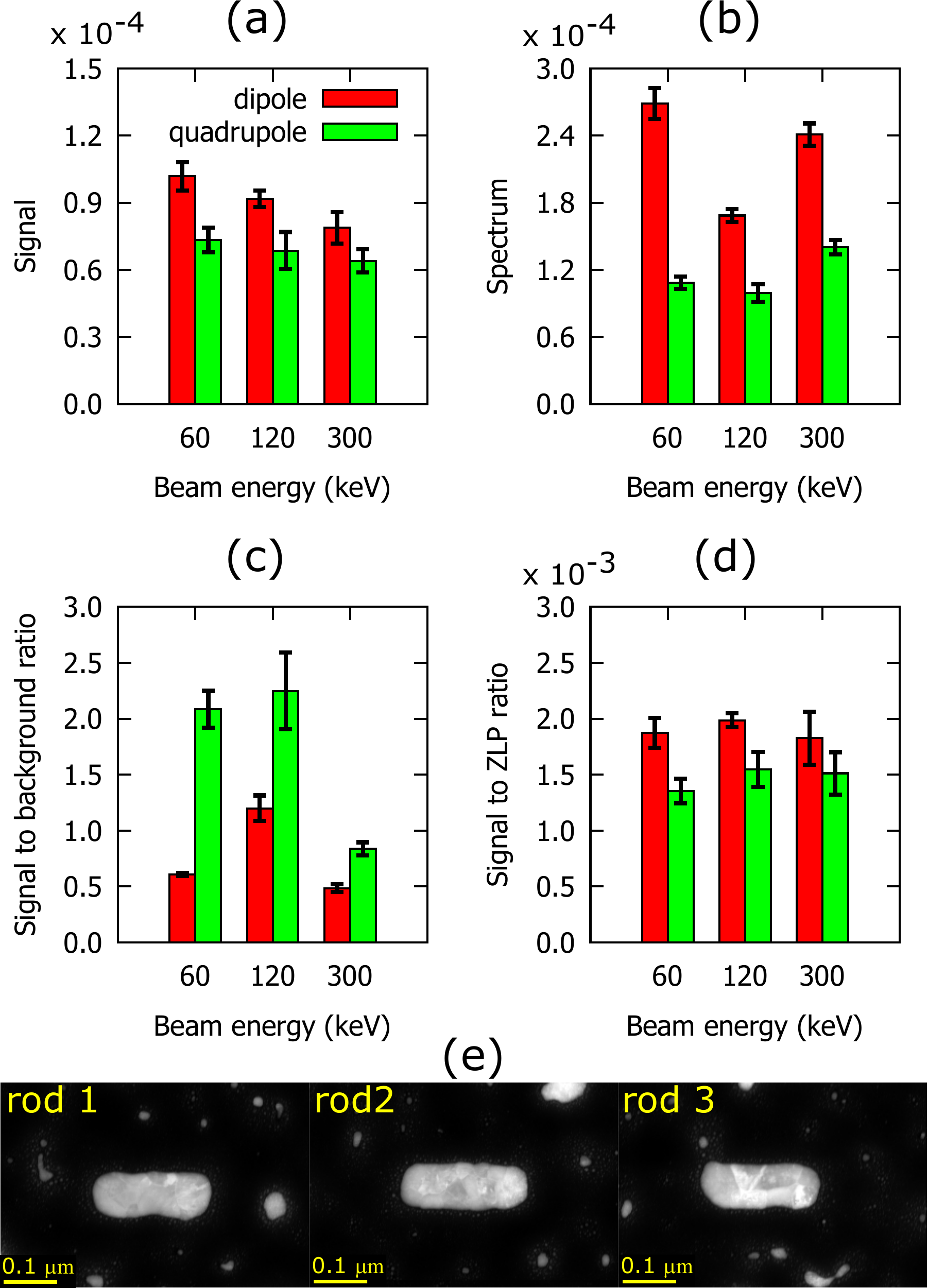}
\caption{Signal (a), spectrum (b), signal to background ratio (c), and signal to ZLP ratio (d) evaluated from a series of 3 rods for longitudinal dipole (red) and quadrupole (green) mode at different primary beam energies. (e) STEM-ADF micrographs of the evaluated series of rods.}
\label{Fig5}
\end{figure}

Up to now, we have evaluated a single rod. To make a general conclusion we have to evaluate the important characteristics --- the signal, the signal to background ratio, and the signal to ZLP ratio --- quantitatively for a series of rods. Figure \ref{Fig5} shows these quantities evaluated from a series of 3 rods [shown in Figure \ref{Fig5}(e)] for longitudinal dipole and quadrupole mode at primary beam energies of 60\,keV, 120\,keV, and 300\,keV. The signal [Figure \ref{Fig5}(a)] is the highest for the primary beam energy of 60\,keV and decreases with increasing the beam energy. The spectrum (i.e., signal plus background) [Figure \ref{Fig5}(b)] is high for the primary beam energies of 60\,keV and 300\,keV and the lowest for the primary beam energy of 120\,keV. The signal to background ratio [Figure \ref{Fig5}(c)] is the highest for the primary beam energy of 120\,keV and the lowest for the primary beam energy of 300\,keV. The signal to ZLP ratio [Figure \ref{Fig5}(d)] is rather constant for all three primary beam energies. Based on these statistics, especially the signal to background ratio, the best results are obtained using 120\,keV electron beam and the worst results are obtained using 300\,keV electron beam.

To conclude, in our case the optimal beam energy to measure LSPR in 30\,nm thick gold rods on a 30\,nm thick silicon nitride membrane by STEM-EELS is 120\,keV. We note that in the case of thicker plasmonic nanostructures using the beam energy of 300\,keV might be advantageous. On the other hand, in the case of a thinner membrane and thinner plasmonic nanostructures using the beam energy of 60\,keV might be useful.

\subsection{Influence of the collection semi-angle on LSPR signal from rods}

\begin{figure}[h!]
\centering
\includegraphics[width=16cm]{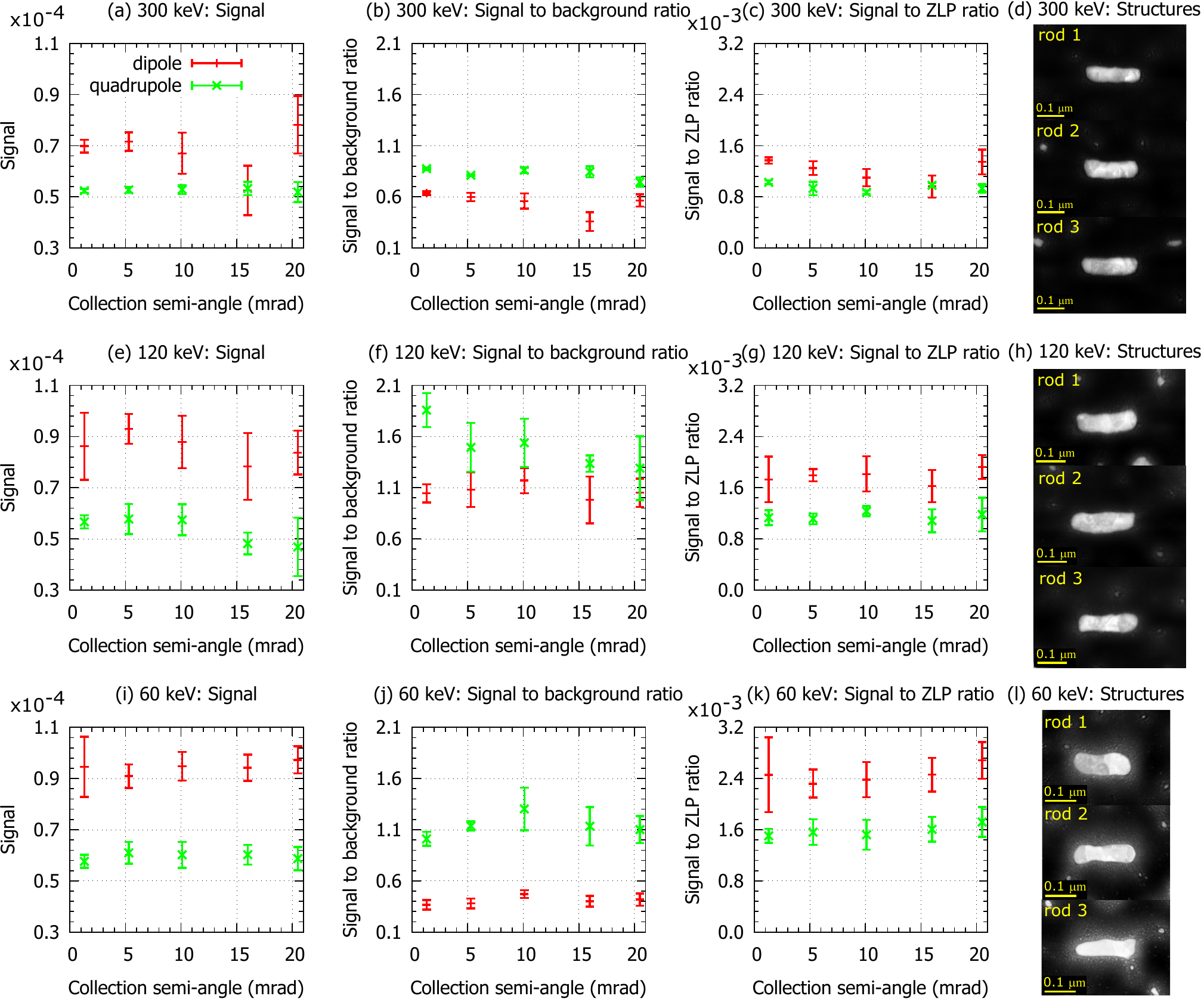}
\caption{Signal (a,e,i), signal to background ratio (b,f,j), and signal to ZLP ratio (c,g,k) evaluated from a series of 3 rods shown in STEM-ADF micrographs (d,h,l) for longitudinal dipole (red) and quadrupole (green) mode as a function of the collection semi-angle at different beam energies (300\,keV, 120\,keV, and 60\,keV). The convergence semi-angle was set to 10\,mrad.}
\label{Fig6}
\end{figure}

In the second experiment we take a series of 3 rods and do the STEM-EELS measurement while changing the the collection semi-angle from 1.3\,mrad to 20.5\,mrad. Note that the convergence semi-angle was constant for all experiments reading 10\,mrad. This experiment was repeated with a different rod series for every primary beam energy, i.e. 300\,keV, 120\,keV, and 60\,keV. The results are summarized in Figure \ref{Fig6}. 

At 300\,keV, the signal [Figure \ref{Fig6}(a)] is maximal for the largest collection semi-angle in the case of the longitudinal dipole mode and rather constant while changing the collection semi-angle for the longitudinal quadrupole mode. The signal to background ratio [Figure \ref{Fig6}(b)] and the signal to ZLP ratio [Figure \ref{Fig6}(c)] reach higher values for either the largest collection semi-angle, or for the smallest collection semi-angle in the case of the longitudinal dipole mode. However, in the case of the longitudinal quadrupole mode, the highest values are reached for the smallest collection semi-angle only. Consequently, at the primary beam energy of 300\,keV the optimal collection semi-angle should be preferentially much smaller than the convergence semi-angle. In our case, we get the best characteristics for the collection semi-angle of 1.3\,mrad. However, we have to notice that the STEM-EELS measurement with the collection semi-angle of 1.3\,mrad is approximately 10times slower as we detect in the spectrometer just a small portion of electrons.

At 120\,keV, the signal [Figure \ref{Fig6}(e)] is higher for smaller collection semi-angles in the case of both modes. The signal to background ratio [Figure \ref{Fig6}(f)] reaches the highest value for the collection semi-angle equal to the convergence semi-angle in the case of the longitudinal dipole mode and for the smallest collection semi-angle in the case of the longitudinal quadrupole mode, respectively. The signal to ZLP ratio [Figure \ref{Fig6}(g)] is rather constant while changing the collection semi-angle for both modes. Consequently, at the primary beam energy of 120\,keV the optimal collection semi-angle should be the same or smaller than the convergence semi-angle. Note that the characteristics are rather flat so at 120\,keV the collection semi-angle is not a critical parameter.

At 60\,keV, the signal [Figure \ref{Fig6}(i)] is rather constant while changing the collection semi-angle for both modes. The signal to background ratio [Figure \ref{Fig6}(j)] reaches the highest value for the collection semi-angle equal to the convergence semi-angle for both modes. The signal to ZLP ratio [Figure \ref{Fig6}(k)] is rather constant while changing the collection semi-angle for both modes. Consequently, at the primary beam energy of 60\,keV the optimal collection semi-angle should be equal to the convergence semi-angle.

To conclude, the collection semi-angle is generally not as critical parameter as the primary beam energy. In the case of a high primary beam energy (represented by 300\,keV), the collection semi-angle should be preferentially much smaller than the convergence semi-angle, but it is not a critical mistake to use the collection semi-angle larger than the convergence semi-angle. In the case of a medium primary beam energy (represented by 120\,keV), the collection semi-angle is not critical at all. Finally, in the case of a low primary beam energy (represented by 60\,keV), the collection semi-angle should be preferentially the same as the convergence semi-angle.

\subsection{Bow-tie and diabolo antennas}

To ensure that the results are valid generally, we have done the experiments with gold bow-tie and diabolo antennas, too. First, we have repeated the primary beam energy experiment on 4 plasmonic antennas. Second, we have repeated the collection semi-angle experiment at the primary beam energy of 300\,keV for one bow-tie antenna. All structures were of a similar size: wing angle of 90$^\circ$ and wing length in range from 230 nm to 250 nm. Such structures support the TD mode around 0.8\,eV and the LD mode around 1.15\,eV.

\begin{figure}[h!]
\centering
\includegraphics[width=8cm]{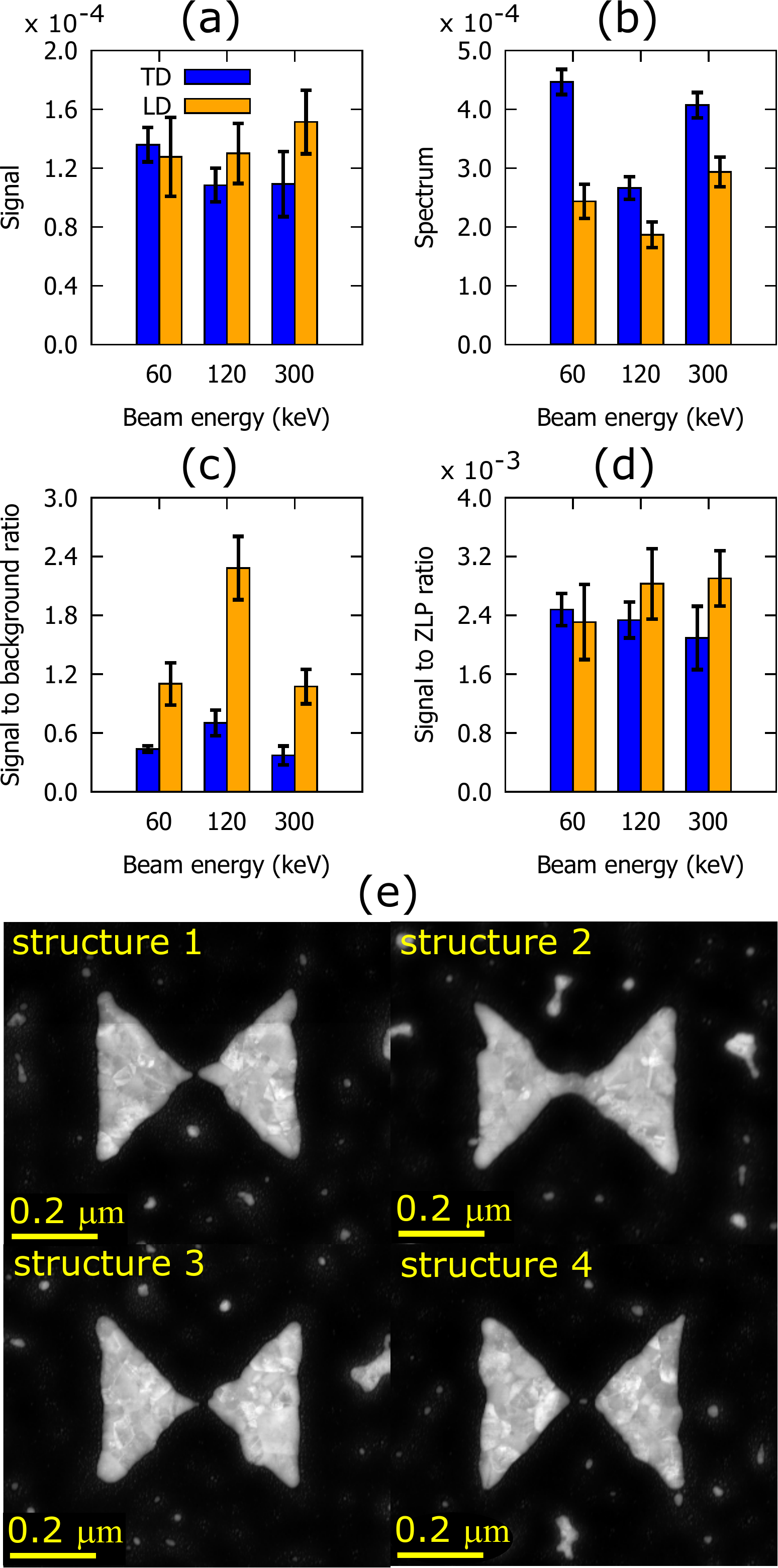}
\caption{Signal (a), spectrum (b), signal to background ratio (c), and signal to ZLP ratio (d) evaluated from a series of 4 bow-tie and diabolo antennas for the TD (blue) and the LD (orange) mode at different primary beam energies. (e) STEM-ADF micrographs of the evaluated series of 3 bow-ties and a diabolo.}
\label{Fig7}
\end{figure}

Figure \ref{Fig7} shows the signal, the spectrum, the signal to background ratio, and the signal to ZLP ratio evaluated from a series of 4 bow-tie and diabolo antennas [shown in Figure \ref{Fig7}(e)] for the TD and the LD mode at primary beam energies of 60\,keV, 120\,keV, and 300\,keV. The signal [Figure \ref{Fig7}(a)] is the highest for the primary beam energy of 60\,keV in the case of TD mode and rather constant in the case of LD mode. The spectrum (i.e. signal plus background) [Figure \ref{Fig7}(b)] is high for the primary beam energies of 60\,keV and 300\,keV and the lowest for the primary beam energy of 120\,keV. The signal to background ratio [Figure \ref{Fig7}(c)] is the highest for the primary beam energy of 120\,keV. The signal to ZLP ratio [Figure \ref{Fig7}(d)] is rather constant for all three primary beam energies. Based on these statistics, especially the signal to background ratio, the best results are obtained using 120\,keV electron beam.

\begin{figure}[h!]
\centering
\includegraphics[width=8cm]{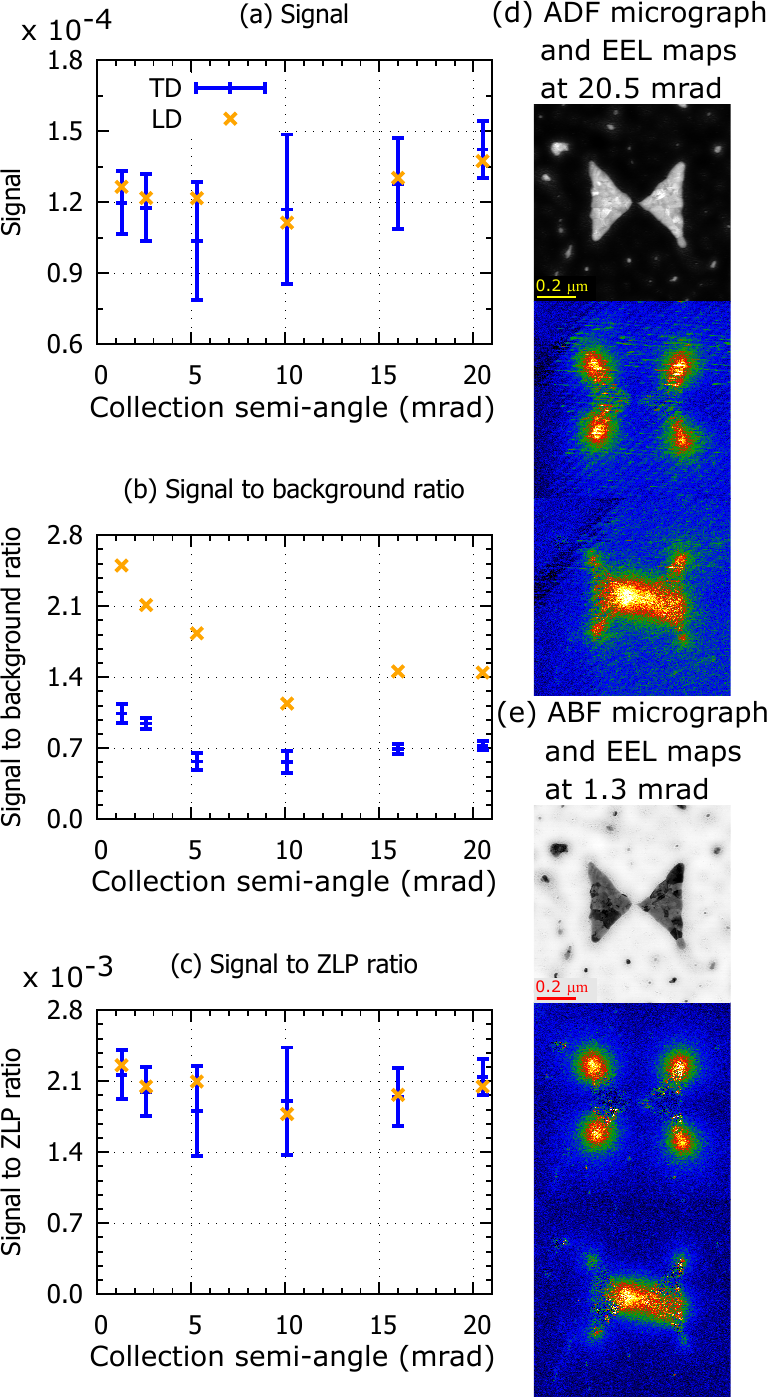}
\caption{Signal (a), signal to background ratio (b), and signal to ZLP ratio (c) evaluated from a bow-tie for the TD (blue) and the LD (orange) mode as a function of the collection semi-angle at the beam energy of 300\,keV. The convergence semi-angle was set to 10\,mrad. Panels (d) and (e) show the STEM micrographs of the bow-tie recorded as an annular dark field (ADF) image for EELS collection semi-angle of 20.5\,mrad (d) and as an annular bright field (ABF) image for EELS collection semi-angle of 1.3\,mrad (e) followed by EEL maps of the TD mode at 0.8\,eV and the LD mode at 1.13\,eV. In the EEL maps the temperature color scale is used, i.e. yellow corresponds to maximal values and blue corresponds to minimal values.}
\label{Fig8}
\end{figure}

Figure \ref{Fig8} shows the signal, the signal to background ratio, and the signal to ZLP ratio evaluated from a bow-tie antenna at 300\,keV primary beam energy while changing the collection semi-angle from 1.3\,mrad to 20.5\,mrad. Note that the convergence semi-angle was constant for all experiments reading 10\,mrad. The signal [Figure \ref{Fig8}(a)] is maximal for the largest collection semi-angle. The signal to background ratio [Figure \ref{Fig8}(b)] and the signal to ZLP ratio [Figure \ref{Fig8}(c)] reach the highest values for the smallest collection semi-angle. Note that relatively high values are reached for the largest collection semi-angle, too. Moreover, there are no significant differences in the EEL maps recorded with the collection semi-angle 1.3\,mrad [Figure \ref{Fig8}(d)] and 20.5\,mrad [Figure \ref{Fig8}(e)]. Consequently, the optimal collection semi-angle should be preferentially much smaller than the convergence semi-angle. In our case, we get the best characteristics for the collection semi-angle of 1.3\,mrad. 

To conclude, the results of the measurement of bow-tie and diabolo antennas have a good agreement with the results obtained for the rods presented in previous sections in Figure \ref{Fig5} and Figure \ref{Fig6}(a-c).

\section{Conclusion}

We have experimentally studied the influence of the primary beam energy and the collection semi-angle on the localized surface plasmon resonances measurement by STEM-EELS. We have discussed the impact on experimental characteristics which are important to detect localized surface plasmon peaks in EELS successfully, namely: the intensity of plasmonic signal, the signal to background ratio, and the signal to zero-loss peak ratio considering a limited dynamic range of the spectrometer camera.

The best results in terms of the best signal to background ratio are obtained using a medium primary beam energy, in our case 120\,keV. In the case of too high primary beam energy, for example 300\,keV, the relativistic effects in the supporting membrane play a non-negligible role and lead to a higher intensity of the background. On the other hand, using too low primary beam energy, for example 60\,keV, leads to a larger relative thickness of the sample so the scattering probability is higher and this results into a higher intensity of the background, too.

The collection semi-angle is generally not as critical parameter as the primary beam energy. Using the primary beam energy of 300\,keV, the collection semi-angle should be preferentially much smaller than the convergence semi-angle, but it is not a critical mistake to use the collection semi-angle larger than the convergence semi-angle. In the case of a medium primary beam energy (represented by 120\,keV), the collection semi-angle is not a critical parameter. Using a low primary beam energy (represented by 60\,keV), the collection semi-angle should be preferentially the same as the convergence semi-angle.

\section*{Acknowledgements}
This work has been supported by the Ministry of Education, Youth and Sports of the Czech Republic under the projects CzechNanoLab [project No. LM2018110, 2020-2022] supporting the CEITEC Nano Research Infrastructure and CEITEC 2020 [project No. LQ1601] and Brno University of Technology [project No. FSI-S-17-4482]. M. H. acknowledges the support of Thermo Fisher Scientific and CSMS scholarship 2019.

\bibliography{bibfile}

\end{document}